\documentclass[journal = jpccck, manuscript = letter]{achemso}
\usepackage{achemso}
\setkeys{acs}{usetitle = true}
\usepackage{graphicx,epsf,epsfig}
\usepackage{dcolumn}
\usepackage{bm}
\usepackage{color}
\usepackage{subfig}
\usepackage{upgreek}
\usepackage{epstopdf}
\usepackage{url}
\usepackage{amsmath}
\usepackage{amssymb}
\usepackage{breakurl}
\usepackage{braket}
\usepackage{lmodern}
\usepackage[T1]{fontenc}
\setkeys{acs}{
  usetitle = true,
  etalmode = truncate,
  maxauthors = 10
}

\definecolor{Green}{RGB}{0,204,102}
\definecolor{Purple}{RGB}{102,0,255}
\definecolor{Blue}{RGB}{51,153,255}
\definecolor{Red}{RGB}{101,010,010}

\title[\texttt{achemso}]
{Energy Pooling Upconversion in Organic Molecular Systems}  

\author{Michael D. LaCount}
\affiliation{Department of Physics, Colorado School of Mines, Golden, CO 80401, USA}
\author{Daniel Weingarten}
\affiliation{Department of Physics, University of Colorado: Boulder, Boulder, CO 80309, USA}
\author{Nan Hu}
\affiliation{Department of Chemistry and Biochemistry, University of Colorado: Boulder, Boulder, CO 80309, USA}
\author{Sean E. Shaheen}
\affiliation{Department of Electrical, Computer, and Energy Engineering, University of Colorado: Boulder, Boulder, CO 80309, USA}
\affiliation{National Renewable Energy Laboratory, Golden, CO 80401, USA}
\author{Jao van de Lagemaat}
\affiliation{National Renewable Energy Laboratory, Golden, CO 80401, USA}
\author{Garry Rumbles}
\affiliation{National Renewable Energy Laboratory, Golden, CO 80401, USA}
\author{David M. Walba}
\affiliation{Department of Chemistry and Biochemistry, University of Colorado: Boulder, Boulder, CO 80309, USA}
\author{Mark T. Lusk}
\email{mlusk@mines.edu}
\phone{(303) 273-3675}
\fax{(303) 273-3919}
\affiliation{Department of Physics, Colorado School of Mines, Golden, CO 80401, USA}

\begin{document}

\begin{abstract}
\noindent A combination of molecular quantum electrodynamics, perturbation theory, and ab initio calculations was used to create a computational methodology capable of estimating the rate of 3-body singlet upconversion in organic molecular assemblies. The approach was applied to quantify the conditions under which such relaxation rates, known as energy pooling, become meaningful for two test systems: stilbene-fluorescein and hexabenzocoronene-oligothiophene. Both exhibit low intra-molecular conversion but inter-molecular configurations exist in which pooling efficiency is at least 90\% when placed in competition with more conventional relaxation pathways.  For stilbene-fluorescein, the results are consistent with data generated in an earlier experimental investigation. Exercising these model systems facilitated the development of a set of design rules for the optimization of energy pooling.
\end{abstract}

\maketitle

\newpage

\section{INTRODUCTION}

The frequency of available light is often lower than desired for a given task. For instance, low-energy light can take advantage of an optical transparency window of biological tissue~\cite{Prasad} and penetrate to ingested theranostic nanoparticles,\cite{Theranostics_Chen_2014} but the photoactivated release of cancer drugs requires energy in the UV range~\cite{Cancer_Dai_2013}.  Within the same backdrop of internal medicine, the fluorescence of biomarkers is difficult to detect if it is in the same frequency as the light used to excited them~\cite{Ohulchanskyy_2010}. The efficiency of photochemical processes in biological systems also suffers as a result of a mismatch of existing and desired frequencies of light. For instance, microalgae used in energy harvesting strategies do not efficiently utilize the red end of the solar spectrum for biomass production~\cite{Biofuel_Upconversion_Menon_2014}. An analogous issue confronts those involved in the design of solar panels, since a large swath of the solar spectrum reaches earth at frequencies too low to be utilized by existing photovoltaic technology~\cite{Xie_2012}. 

\begin{figure}[h]
\centering{}
\subfloat[ ]{\includegraphics[width=0.19\textwidth]{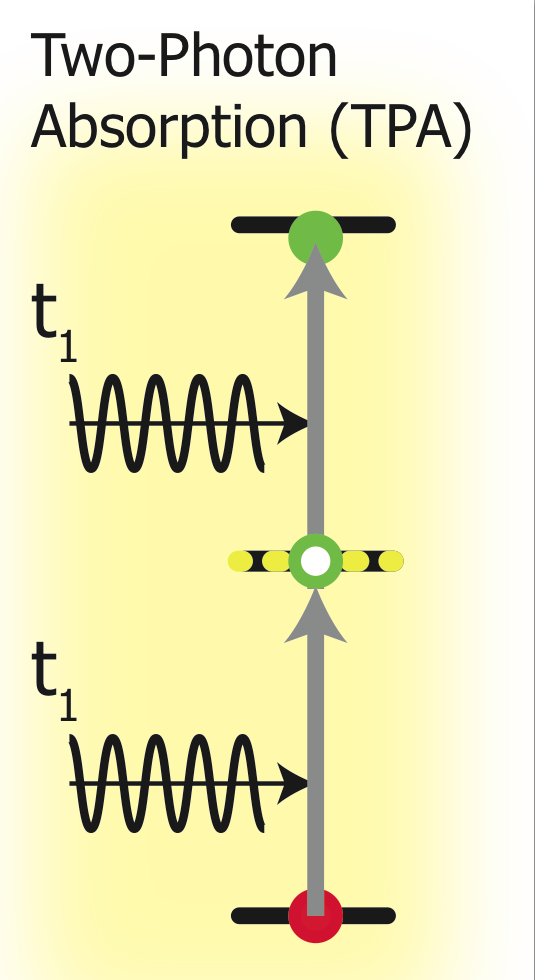}}
\qquad
\subfloat[ ]{\includegraphics[width=0.19\textwidth]{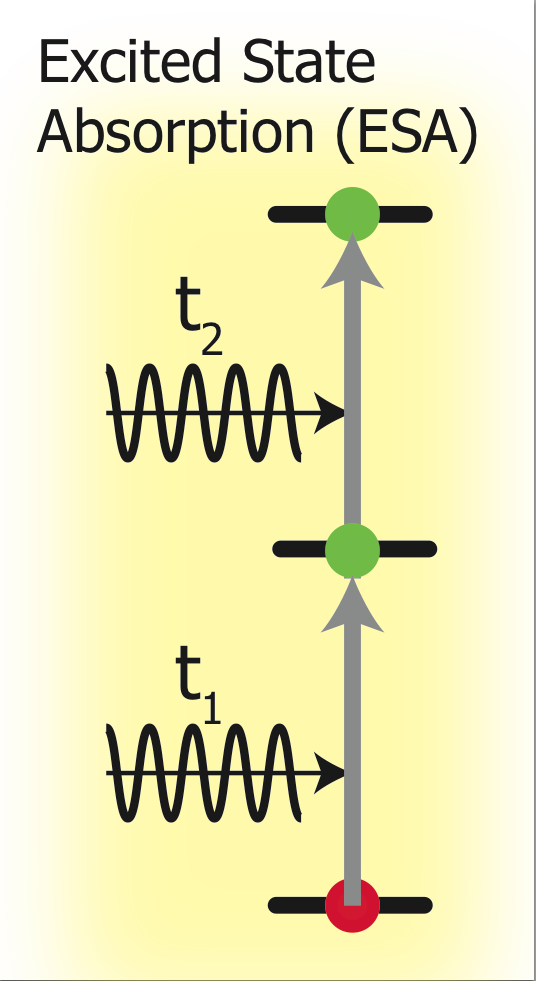}}
\caption{ Energy level diagrams for two-photon absorption (a) and excited state absorption (b). Times are indicated by $t_1$ and $t_2$. Solid black lines indicate eigenstates. Yellow-black lines indicate virtual states.}
\label{TPA_ESA}
\end{figure}

Motivated by such disconnects between the desired light and that available, a number of strategies have been developed to upconvert light to higher frequencies. One of the simplest upconversion methods is the two-photon absorption (TPA) shown at left in Fig. \ref{TPA_ESA}, but this occurs only under extremely high intensity light.\cite{Pawlicki_2009, Bartolo_2011}  Nanostructured condensed matter, though, can be fabricated to independently absorb low-energy photons and subsequently combine the separate excitons into a single, higher energy state. Such upconversion processes are distinct from nonlinear optics in that the initial excitons can be created at distinctly separate times and, typically, on differing nanostructures. These processes, for instance, allow the infrared spectrum to be utilized to carry out high-energy tasks in photovoltaics\citep{Xie_2012,Zou_2012,Trupke_2006}, biofuel production and medical applications.\citep{Ang_2011,Chatterjee_2010}

\begin{figure}[h]
\centering{}
\includegraphics[width=0.70\textwidth]{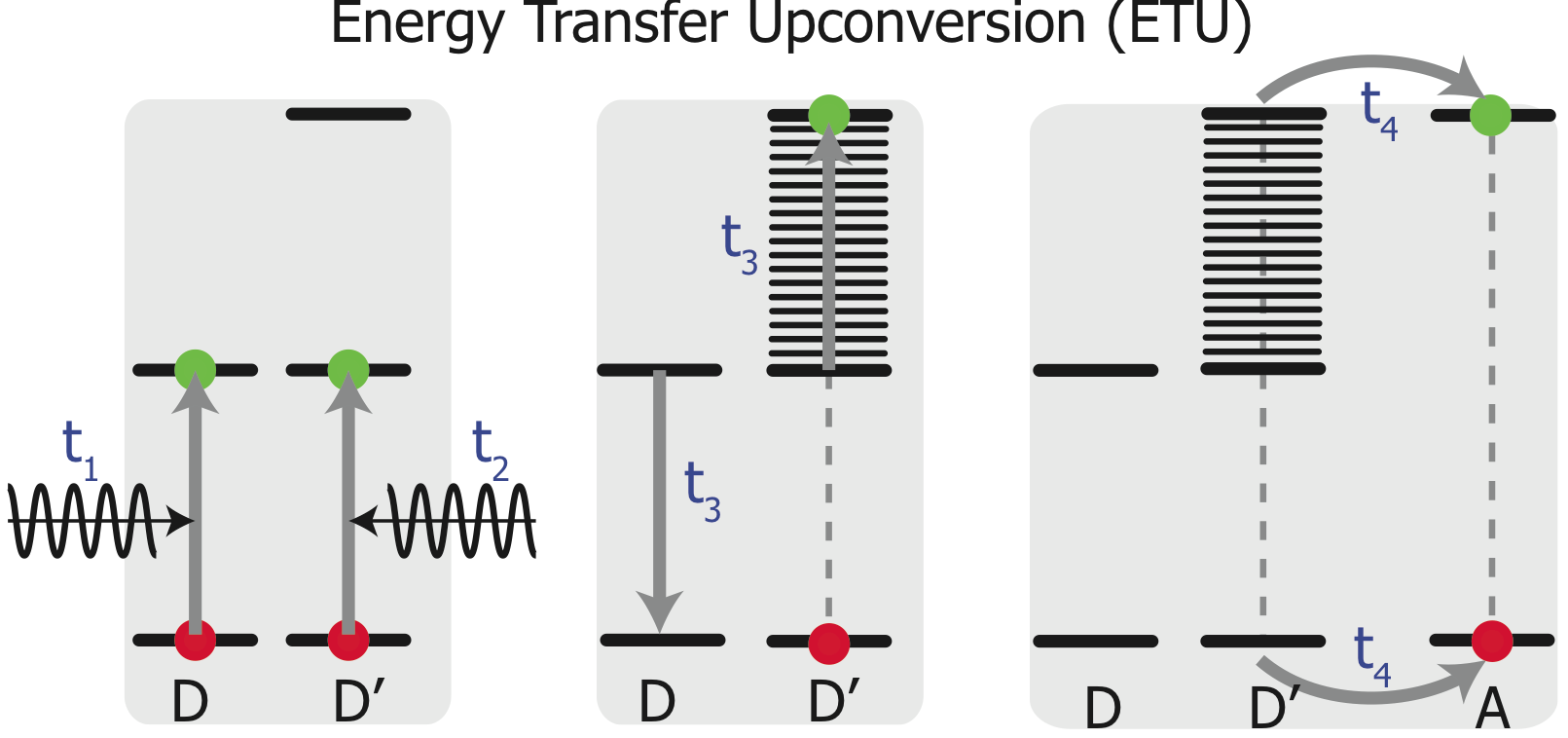}
\caption{ Energy level diagrams for two-electron energy transfer upconversion (ETU) from donor molecules (D and D') to an acceptor molecule (A). Solid black lines indicate eigenstates. Times are indicated by $t_1$--$t_4$. }
\label{ETU}
\end{figure}

In principle, upconversion can be carried out by sequentially exciting a single electron, known as Excited State Absorption (ESA), as shown at right in Fig. \ref{TPA_ESA}. However, it is difficult to design systems in which ESA is efficient while ensuring that the resulting energy can be transferred before the exciton thermalizes.\cite{Theranostics_Chen_2014} Much more practical to implement is a two-electron process, shown in Fig. \ref{ETU}, in which a sensitizer molecule absorbs photons and transfers excitons to an activator. The latter are typically lanthanide atoms for which the 4f excitation is encapsulated within the 5s bonding orbital.\citep{Wang09} A low electron-phonon coupling results and the thermalization is consequently slow.

%

From the perspectives of biocompatibility, environmental safety, expense and resource abundance, it would be advantageous to use organic materials for such Energy Transfer Upconversions (ETU). High reorganization energy, and its manifestation as rapid phonon generation, has precluded any meaningful organic molecular ETU, though, because excited states thermalize into the first excited state faster than any other competing process.\cite{Theranostics_Chen_2014} It was this issue, in fact, that drove research towards lanthanide nanoparticles in the first place. This has motivated a consideration of three-electron processes\cite{Jenkins} that completely side-step the thermalization issue. Energy Pooling (EP) is defined as a three electron process in which the energy of two electrons in excited states are transfered to the third electron. Such EP dynamics access virtual states to mediate upconversion, so there is no residence time during which an excited state has the chance to lose energy to phonons. 

%

\begin{figure}[h]
\centering{}
\subfloat[ ]{\includegraphics[width=0.40\textwidth]{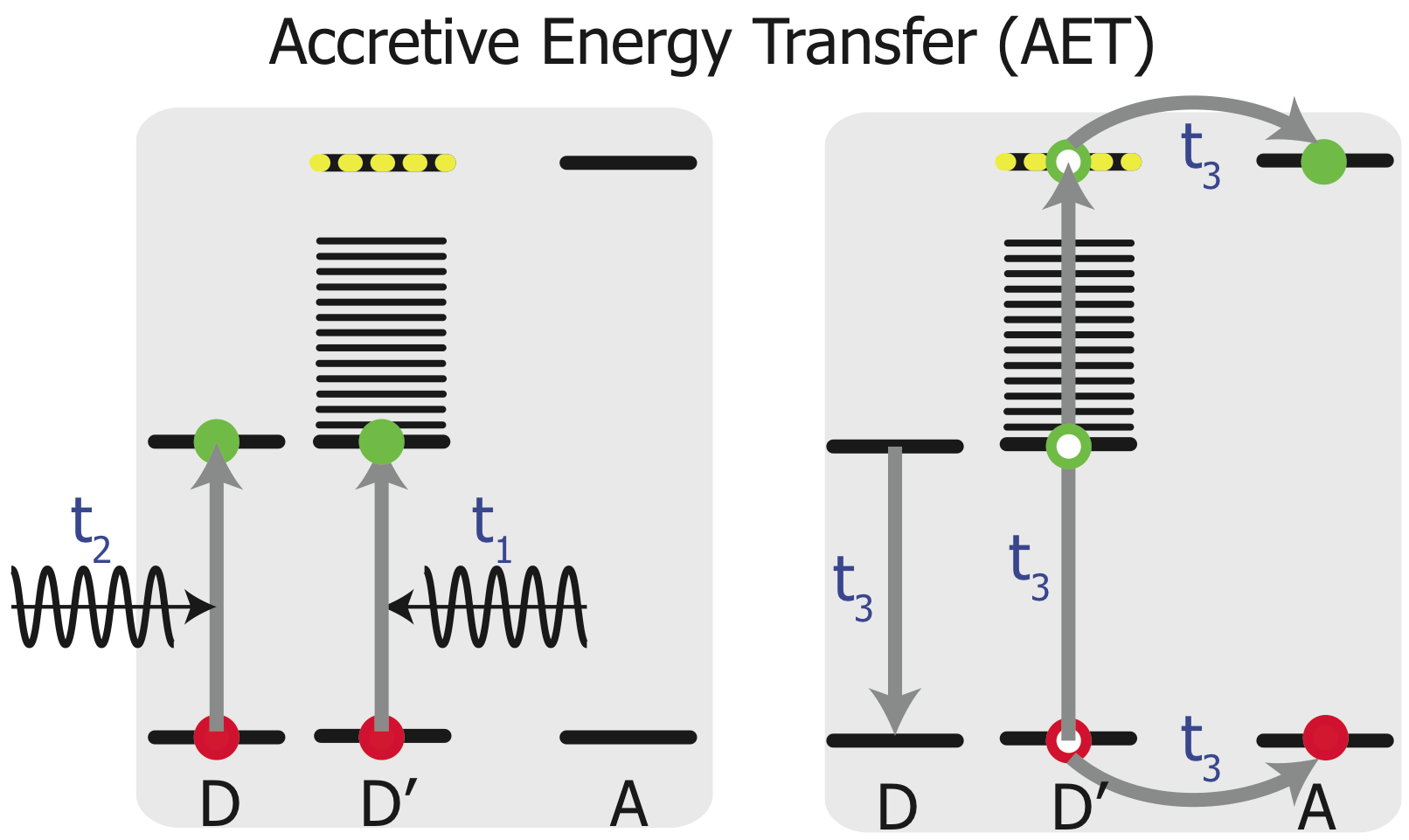}}
\qquad
\subfloat[ ]{\includegraphics[width=0.40\textwidth]{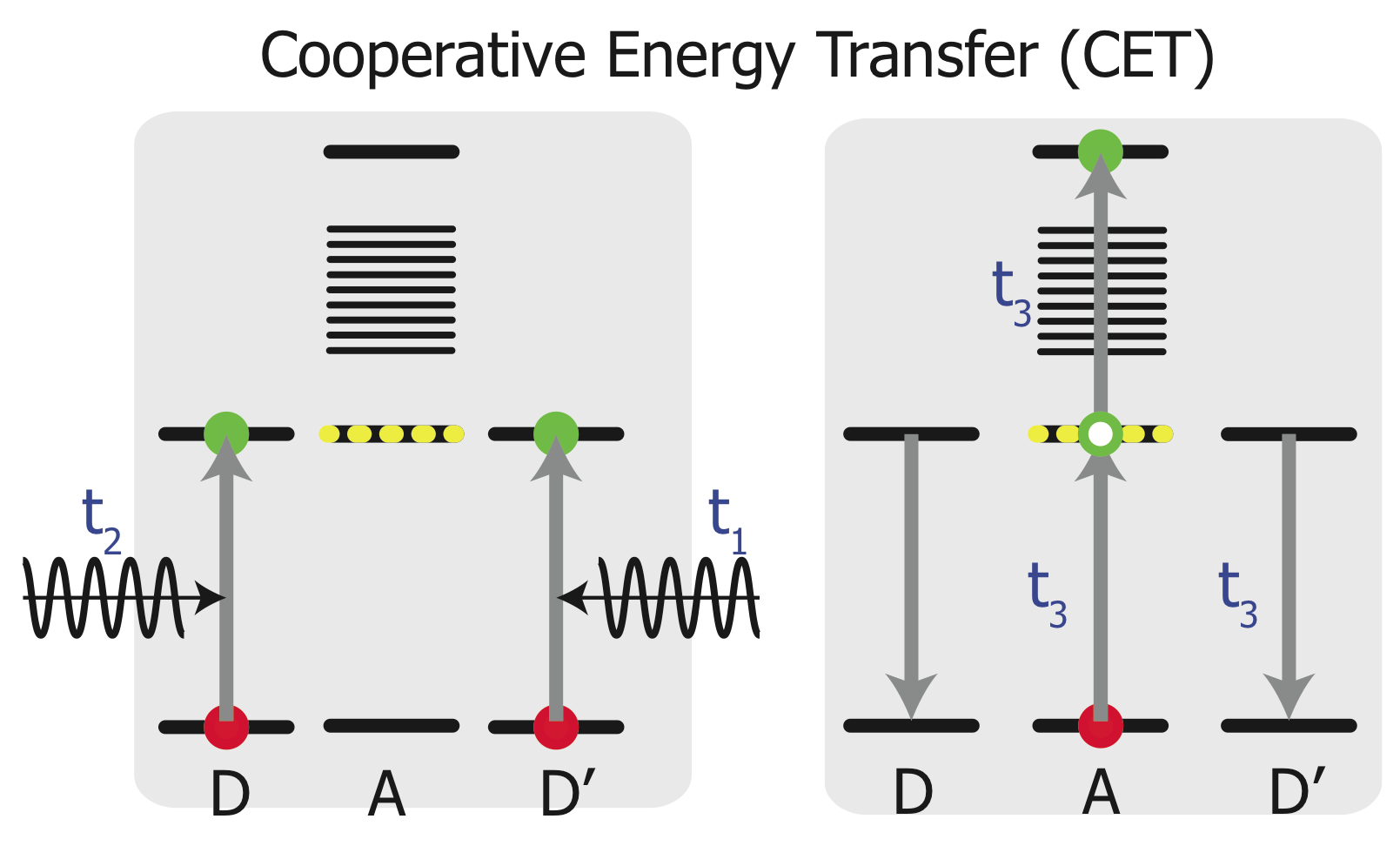}}
\caption{Schematic depiction of two energy pooling (EP) mechanisms from donor molecules (D and D') to an acceptor molecule (A). Solid black lines indicate eigenstates. Times are indicated by $t_1$--$t_3$. Yellow-black lines indicate virtual states.}
\label{processes}
\end{figure}

EP can occur via the two separate mechanisms illustrated in Fig. \ref{processes}. The first is Accretive Energy Transfer (AET) in which one donor transfers its energy to a second donor, placing that donor into a virtual state. The second donor transfers its energy to the acceptor, and all time orderings contribute to the dynamics. The second pooling process is Cooperative Energy Transfer (CET) in which the energy of one donor is transferred to the acceptor placing it into a virtual state. A second donor also transmits its energy to the acceptor, and the combined result is that the acceptor is in one of its eigenstates. As with AET, all time ordering contribute the the total pooling rate. 
%
Although experimentally observed for lanthanide nanoparticles,\cite{Liang_2009, Pushkar_2011} the conditions under which EP will occur in organic condensed matter has not been explored. An earlier experimental study of a stilbene-fluorescein molecule~\citep{stilbenefluorescein}  suggests the possibility of organic upconversion providing an initial focus for the current computational investigation. Upconversion was observed to have occurred under high intensity excitation of the fluorescein donor resulting in excitation of the stilbene core. It was observed that unless two donors were bonded to each stilbene no upconversion occurred, eliminating TPA as a possible explanation. The paper concludes that either triplet-triplet annihilation (TTA) upconversion or some three-body resonance energy transfer (EP) could be responsible for the observed data but leans towards TTA.

\begin{figure}[h]
\begin{centering}
\includegraphics[height=1.5in]{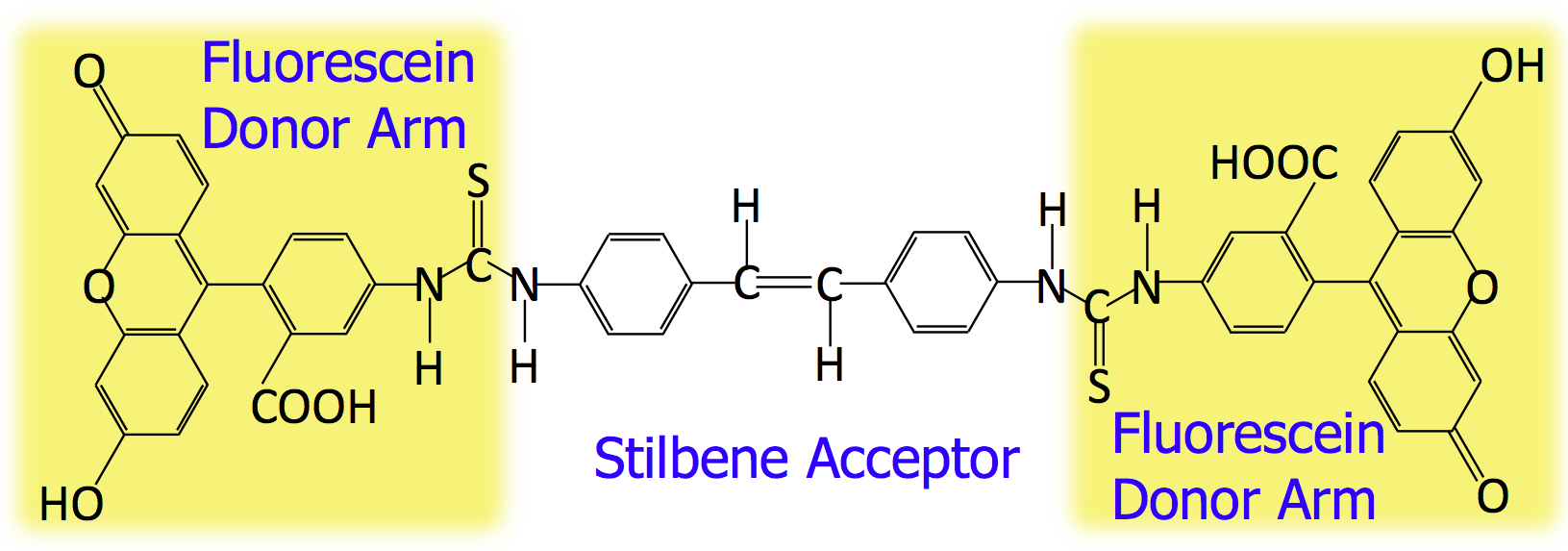}
\par\end{centering}
\caption{E-4,4$'$-Di(fluorescein-5$"$-yl-thioureanyl)stilbene with two fluorescein donor arms connected to a stilbene acceptor.}
\label{Fluorescein_Stilbene}
\end{figure}

\begin{figure}
\begin{centering}
\includegraphics[height=3.5in]{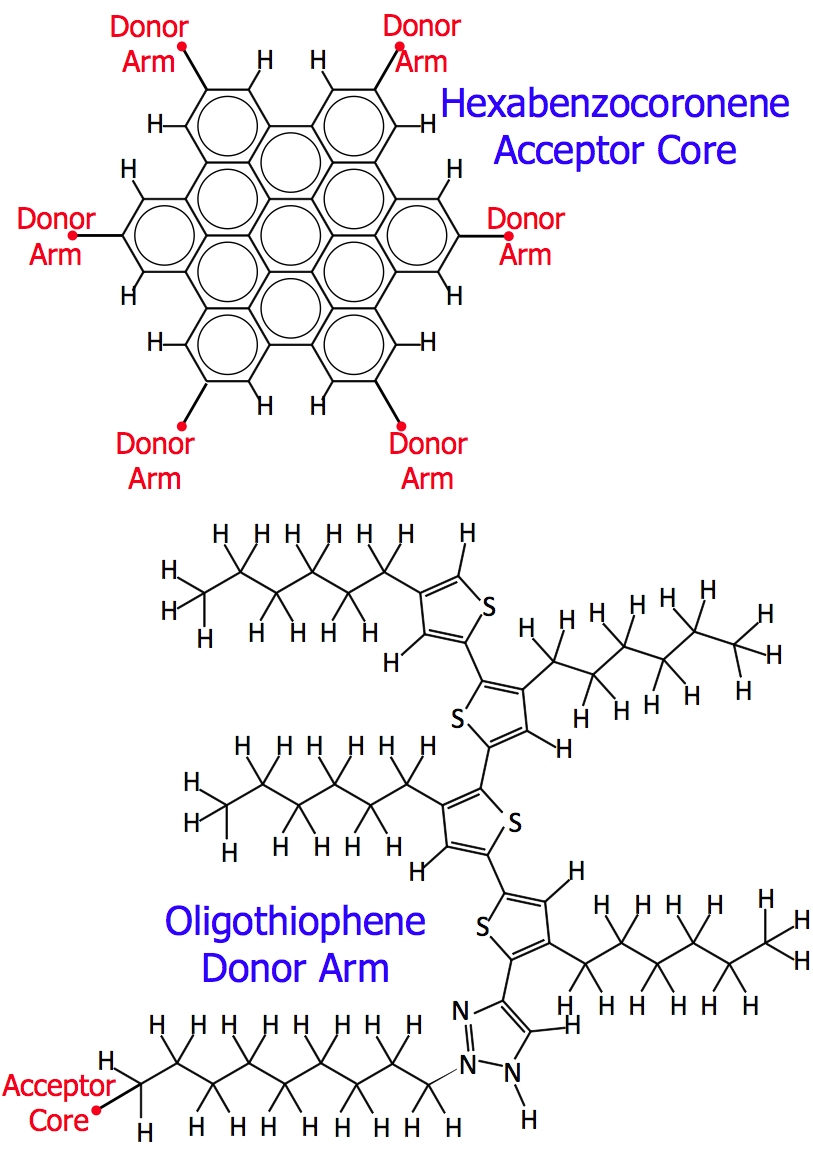}
\par\end{centering}
\caption{Hexabenzocoronene (HBC) acceptor at symmetrically connected to six oligothiophene (OTP) donor arms.}
\label{HBC_OTP}
\end{figure}

A quantum electrodynamics (QED) approach is combined with ab initio excited state structure calculations to quantify upconversion rates in assemblies of organic molecules. The codes are applied in a parametric study to formulate a set of design rules that can be used to guide the synthesis of systems that can efficiently carry out energy pooling. The computational suite is subsequently applied to quantify upconversion processes in two condensed matter assemblies: stilbene-fluorescein molecules as shown in Fig. \ref{Fluorescein_Stilbene}; and hexabenzocoronene(HBC)-oligothiophene(OTP) molecules as shown in Fig. \ref{HBC_OTP}. 

The stilbene-fluorescein molecule studied is shown in Fig. \ref{Fluorescein_Stilbene}. It is assumed that the molecule can be divided into a 4,4$'$-diaminostilbene core and fluorescein-5-isothiocyanate antennae components and each may be studied independently of each other. The interaction between the core and antennae are assumed to be dominated by dipole coupling.
%
%

%

\section{METHODOLOGY}

\begin{figure}[h]
\begin{centering}
\includegraphics[width=4.5in]{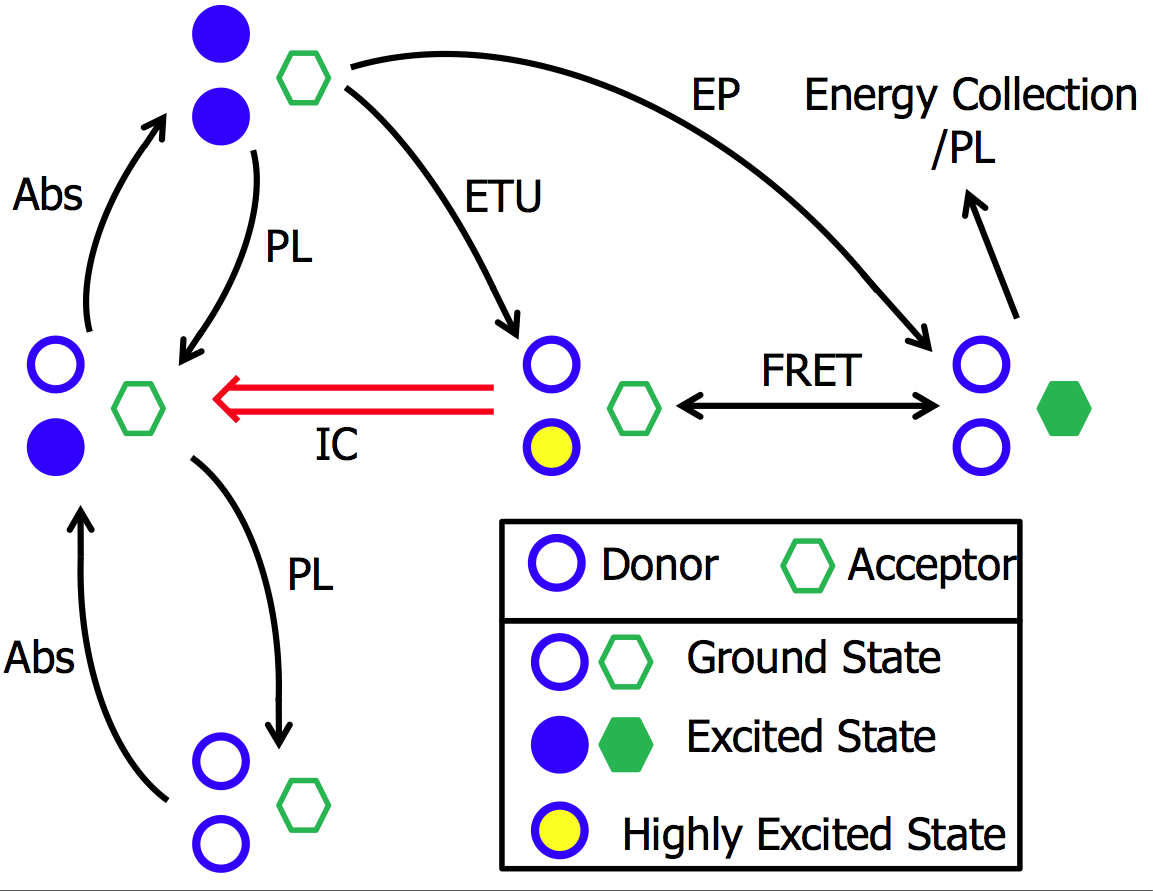}
\par\end{centering}
\caption{Schematic of energy transfer pathways considered in the computational paradigm: absorption (Abs), photoluminescence (PL), Energy Transfer Upconversion (ETU), Energy Pooling (EP), and F\"{o}rster Resonance Energy Transfer (FRET). Internal conversion (IC), highlighted in red, is assumed to occur much faster than all other processes.}
\label{pathways}
\end{figure}

Molecular QED provides a description of the competing energy transfer pathways, and those considered are shown in Fig. \ref{pathways}. Internal conversion and thermalization were not considered explicitly. Instead, it was assumed that electron-phonon coupling is sufficiently high that such processes, when available, dominate any relaxation dynamics. For instance, transitions to levels above the first excited state are assumed to immediately cool to the first excited state.
%

\begin{figure}[h]
\begin{centering}
\includegraphics[width=4.5in]{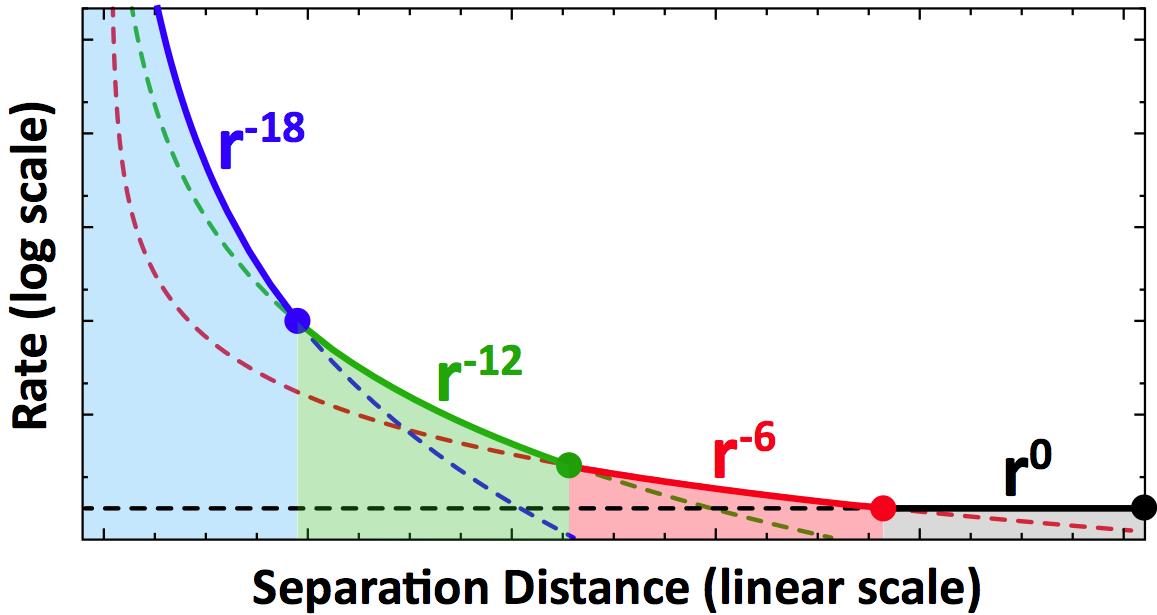}
\par\end{centering}
\caption{Idealization of perturbation processes of several orders as a function of molecular separation. The r$^{0}$, r$^{-6}$, r$^{-12}$, and r$^{-18}$ curves are scaled by factors of 1, 0.2, 2$\times$10$^{-3}$, and 10$^{-7}$, respectively.}
\label{PTrates}
\end{figure}

All relaxation rates between initial and final states, $\Ket{i}$ and $\Ket{f}$, were derived from $n^{\rm th}$-order perturbation theory and so are expressed in terms of a coupling matrix element, $W^{(n)}$, and an energy density of states, $\delta$:
\begin{equation}
\Gamma_{fi}=\frac{2\pi}{\hbar}\left|\Bra{f}W^{(n)}\Ket{i}\right|^{2}\delta(E_{f}-E_{i})
\label{eq:FGR}
\end{equation}
In turn, the coupling, W$^{(n)}$, has a functional dependence on the interaction Hamiltonian, $V'$, that depends on the order of perturbation theory, $n$:
\begin{eqnarray}
W^{(1)}&=&V' \label{eq:pertorder1}\\
W^{(2)}&=&\sum_{r\neq i}\frac{V'\Ket{r}\Bra{r}V'}{E_{i}-E_{r}} \label{eq:pertorder2} \\
W^{(4)}&=&\sum_{r,s,t\neq i}\frac{V'\Ket{r}\Bra{r}V'\Ket{s}\Bra{s}V'\Ket{t}\Bra{t}V'}{(E_{i}-E_{r})(E_{i}-E_{s})(E_{i}-E_{t})}
\label{eq:pertorder4}
\end{eqnarray}
It is often the case that the rates of two relaxation processes can be quantified using perturbation theory with each governed by the lowest non-zero interaction matrix element from Eqs. \ref{eq:pertorder1}--\ref{eq:pertorder4}. In such situations, it might be tempting to conclude that the process of lowest order will be the fastest, but this is not always the case since molecular separation must also be taken into account. For instance,  F\"{o}rster Resonance Energy Transfer (FRET) can be more rapid than photoluminescence (PL) even though it is a second order perturbative process while PL is first order. Generally speaking, perturbative dynamics tend to have inverse polynomial relationships with separation distance, $r$, and the $r$-dependence of first, second, fourth, and sixth order processes are r$^{0}$, r$^{-6}$, r$^{-12}$, and r$^{-18}$ dependence, respectively. Fig. \ref{PTrates} shows that it is possible for each process to dominate relaxation over a range of separation distance, with closer distances associated with higher order processes in general. Scaling factors for the r$^{0}$, r$^{-6}$, r$^{-12}$, and r$^{-18}$ curves were chosen to be 1, 0.2, 2$\times$10$^{-3}$, and 10$^{-7}$, respectively. This choice was based on the assumption that the scaling factor should decrease exponentially with increasing order of perturbation theory. This suggests that energy pooling, a fourth-order perturbative process, will be observable only when molecular separations are very small.  
%

For a given relaxation process, the interaction Hamiltonian can be expressed using the electric dipole approximation, and the form is the same for each of type of dynamics considered in this work:
\begin{equation}
V'=-q\mathbf{r}\cdot\mathbf{E}(\mathbf{r})
\end{equation}
\begin{equation}
\mathbf{E}(\mathbf{r})=\imath\sum\limits _{\mathbf{k',\lambda'}}E_{k'}[\hat{a}(\mathbf{k'}\lambda')\vec{\xi}(\mathbf{k'}\lambda')e^{\imath(\mathbf{k'}\cdot\mathbf{r})}-\hat{a}^{\dagger}(\mathbf{k'}\lambda')\vec{\xi'}(\mathbf{k'}\lambda')e^{-\imath(\mathbf{k'}\cdot\mathbf{r})}],\: E_{k'}=\sqrt{\frac{\hbar ck}{2\epsilon_{0}\Omega}}
\end{equation}
In these equations, $\mathbf{k}$ is the wave vector, $\lambda$ is the polarization index, $\hat{a}^{\dagger}(\mathbf{k}\lambda)$ and $\hat{a}(\mathbf{k}\lambda)$ are photon creation and annihilation operators, $\vec{\xi}(\mathbf{k}\lambda)$ and $\vec{\xi'}(\mathbf{k}\lambda)$ are the unit polarization vectors for each photon mode, $c$ is the speed of light in vacuum,  $\epsilon_0$ is the free space dielectric constant, and $\Omega$ is the normalization volume. The rates of photoluminescence (PL), F\"{o}rster Resonance Energy Transfer (FRET), and both pooling processes have previously been simplified into equations whose terms can be measured experimentally or calculated computationally:\cite{PLRate,Jenkins2body,Jenkins,Jenkins1999}\\
\begin{equation}
\Gamma_{\mathrm{PL}}=\frac{k^{3}(\mu^{0\alpha(D)}){}^{2}}{3\pi\varepsilon_{0}\hbar}=\frac{(E_{\mathrm{i}}-E_{\mathrm{f}})^{3}(\mu^{0\alpha(D)}){}^{2}}{3\pi\varepsilon_{0}\hbar^{4}c^{3}}
\end{equation}
\begin{equation}
\Gamma_{\mathrm{FRET}}=\sum_{k,l=1}^3\frac{2\pi}{\hbar}\left|\mu_{k}^{\alpha_{\mathrm{f}}\alpha_{\mathrm{i}}(D)}V_{kl}(k,\mathbf{R})\mu_{l}^{\beta_{\mathrm{f}}\beta_{\mathrm{i}}(A)}\right|^{2}\delta(E_{\mathrm{f}}-E_{\mathrm{i}})
\end{equation}
\begin{equation}
\Gamma_{\mathrm{AET}}=\sum_{k,l,m,n=1}^3\frac{2\pi}{\hbar}\left|\mu_{k}^{0\alpha(D)}V_{kl}(k,\mathbf{R''})\alpha_{lm}^{0\alpha(D')}(2k,-k)V_{mn}(2k,\mathbf{R'})\mu_{n}^{\beta0(A)}\right|^{2}\delta(E_{\mathrm{f}}-E_{\mathrm{i}})
\end{equation}
\begin{equation}
\Gamma_{\mathrm{CET}}=\sum_{k,l,m,n=1}^3\frac{2\pi}{\hbar}\left|\mu_{k}^{0\alpha(D)}V_{kl}(k,\mathbf{R})\alpha_{lm}^{\beta0(A)}(-k,-k)V_{mn}(k,\mathbf{R'})\mu_{n}^{0\alpha(D')}\right|^{2}\delta(E_{\mathrm{f}}-E_{\mathrm{i}})
\end{equation}
with the electric-dipole coupling tensor, $V$ and two-photon absorption tensor, $\alpha$, given by
\begin{equation}
V_{mn}(k,\mathbf{R})=\frac{e^{\mp \imath kR}}{4\pi{}^{2}\varepsilon_{0}R^{3}}((1\pm \imath kR)(\delta_{mn}-3\hat{R_{m}}\hat{R_{n}})-k^{2}R^{2}(\delta_{mn}-\hat{R_{m}}\hat{R_{n}}))
\end{equation}
\begin{equation}
\alpha_{mn}^{\mathrm{f}\mathrm{i}(\xi)}(\pm k_1,\pm k_2)=\sum_{\zeta}\left(\frac{\mu_m^{\mathrm{f}\zeta(\xi)}\mu_n^{\zeta\mathrm{i}(\xi)}}{E_{\mathrm{i}\zeta}+\imath\gamma\mp\hbar c k_1}+\frac{\mu_n^{\mathrm{f}\zeta(\xi)}\mu_m^{\zeta\mathrm{i}(\xi)}}{E_{\mathrm{i}\zeta}+\imath\gamma\mp\hbar c k_2}\right)
\label{eq:tpatensor}
\end{equation}
The following notation is used in these equations:\cite{Jenkins}  
\begin{itemize}
\item{k refers to the magnitude of the wave vector associated with a photon emitted from an excited donor;}
\item{ $\mu^{xy(z)}$ represents the transition dipole moment from state y to state x on molecule z;}
\item{${\bf R}$, ${\bf R}'$, and ${\bf R}''$ are the displacement vectors between molecules D and A, molecules D' and A, and molecules D and D' respectively. Hats indicate unit vectors while ${R}$, ${R}'$, and ${R}''$ are vector magnitudes.}
\item{$\gamma$ is a generic broadening taken to have a value of k$_B$T ($\sim$25 meV). Varying this parameter up to a maximum value of 100 meV caused less than an order of magnitude change in the rates. While this caused a quantitative change in the rates and efficiencies, there was no qualitative change in our results. This is consistent with other studies on the effect the broadening term on energy transfer processes between molecular systems.}\citep{Lin}
\item{$\zeta$ represents the possible virtual states that occur between the initial and final state.}
\end{itemize}
 
It is worth noting that an approximation is sometimes made to simplify the numerical evaluation of the two-photon absorption tensor, $\alpha$, by reducing the sum over all states in Eq. \ref{eq:tpatensor} to the sum of only the first and final state. However, this approximation is only reasonable under very specific conditions: (1) the molecule must posses a electric dipole moment that is larger than its transition dipole moments; (2) no singlet states must be closer in energy to the virtual state than the initial and final state; and (3) neither the initial or final state can have any excited states that are close in energy. The second condition makes this approximation invalid for the AET.  For CET, this second condition can sometimes be met. However, for the acceptor molecules considered in this paper the electric dipole moment is essentially zero for CET so the first condition is not met. 

The square of the two-photon absorption tensor, $\alpha_{ij}\alpha_{kl}$, is proportional to the rate of two-photon absorption, $\Gamma_{TPA}$, given here in three different representations because each has a discipline in which it is commonly used: 
\begin{equation}
\Gamma_{TPA}=\frac{I^2}{2(\hbar c k)}\frac{\beta}{N/\Omega}=\frac{I^2}{2(\hbar c k)^2}\sigma^{(2)}=\frac{3 I^2}{4 n^2 \epsilon_0 \hbar c^2}\frac{{\rm Im}[\chi^{(3)}]}{N/\Omega}
\label{eq:tpaRelationships}
\end{equation}
Here $\beta$ is the two-photon absorption coefficient, $\sigma^{(2)}$ is the two-photon absorption cross-section, and the imaginary part of the third-order electric susceptibility is $\chi^{(3)}$.\cite{Boyd, optics,TPARate2}  In addition, $I$ is the irradiance of the light source and $N$ is the number of molecules in the absorbing material.

Summations are over the entire electronic eigenbasis although only a cutoff is imposed in practice. The energy terms, $E$, refer to the electronic energy only and do not include the energy of photons that may be present in either the initial or final state. A Lorentzian function is used for the electronic density of states:
\begin{equation}
\delta(\Delta E)=\frac{1}{\pi}\frac{\gamma /2}{(\gamma/2)^2+\Delta E^2}\quad .
\end{equation}
 
The set of rate equations summarized above was populated using electronic structure analyses for the stilbene-fluorescein and hexabenzocoronene-oligothiophene geometries shown in Figs. \ref{Fluorescein_Stilbene} and \ref{HBC_OTP}. In the original experimental report on upconversion in stilbene-fluorescein, intermolecular structural information was absent except for a brief comment describing the material as a "red crystalline phase".\citep{stilbenefluorescein} Because of this lack of structural information, the material was treated here as an assembly of interacting molecules without translational symmetry. 

Molecular geometries were calculated using Density Functional Theory (DFT) and Time-Domain (TD)-DFT theory for ground states and excited states, respectively. The Def2-TZVP atomic basis set was used for all calculations.\citep{Basis} However, a series of exchange-correlation functionals were considered that include: PBE, PBE0, B3LYP, and CAM-B3LYP.\citep{PBE,PBE0,B3LYP,CAMB3LYP} The impact on excited state properties of implementing the Tamm-Dancoff approximation (TDA) was also studied.\citep{Hirata1999} Absorption and emission spectra were produced directly from the TD-DFT calculation of the ground state geometry and excited state geometry respectively.\citep{Rohlfing} For comparisons with experimental emission spectra, Kasha's rule was enforced.\citep{Kasha_1950}

Despite being referred to as {\it ab initio} calculations, an entire biom of density functionals have evolved for capturing electron exchange and correlation effects across the spectrum of condensed matter systems. Care must be taken in choosing a functional best suited to the system being studied, and this is particularly true for energy pooling. For instance, discrepancies in donor energy levels are doubly counted in the associated electronic density of states, and products of four transition dipole matrix elements make pooling rates especially sensitive to errors in their estimation. As a guide for choosing exchange/correlation functionals, experimental absorption and emission spectra were measured and compared with those generated using PBE0, CAM-B3LYP, and B3LYP functionals.~\citep{PBE,PBE0,B3LYP,CAMB3LYP}  

UV-vis absorption spectra for HBC was performed in solution using chloroform as a solvent. The emission data for oligothiophene was taken in a solution of chloroform. The oligothiophene samples were placed in 1 cm x 1 cm quartz cuvettes and excited with 380 nm light. Emission data was taken over the range of 390 to 690 nm, measured 90 degrees relative to the excitation. These measurements were performed at room temperature using a PTI fluorometer with an Ushio UXL-75XE xenon short arc lamp and a Hamamatsu R928P PMT tube operating at \textasciitilde 1000 V, DC.

\begin{figure}[h]
\centering{}
\subfloat[Fluorescein-5-Isothiocyanate]{\includegraphics[width=3in]{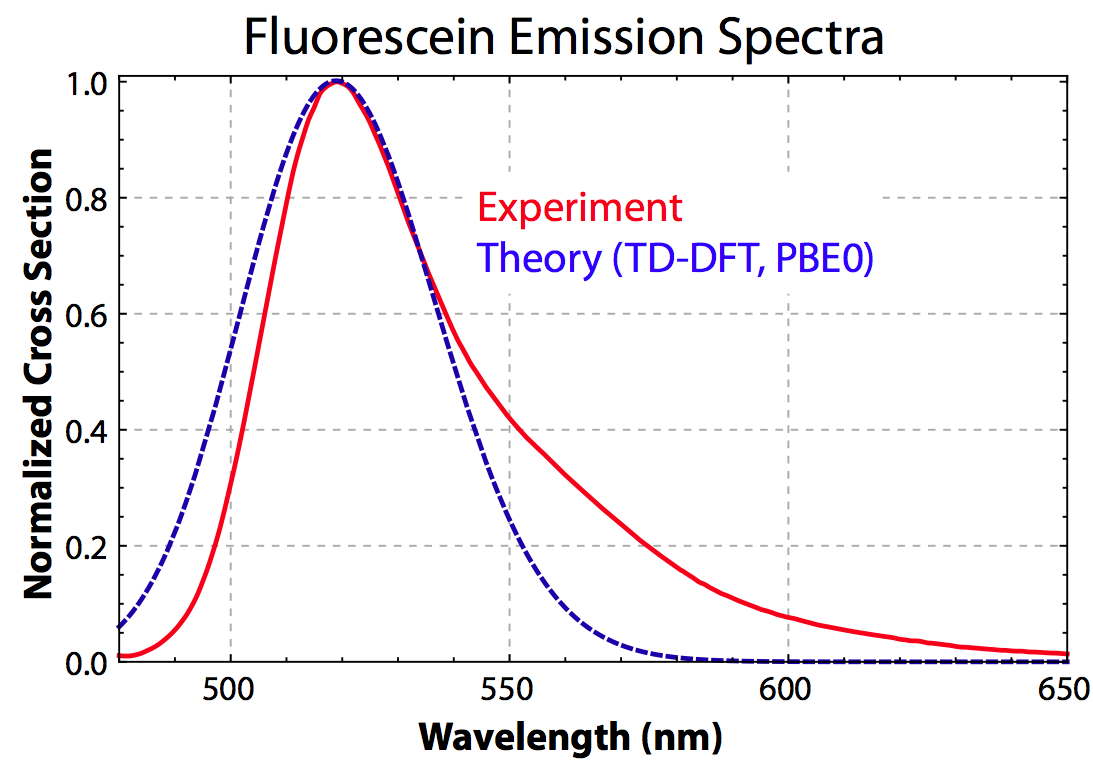}}
\qquad
\subfloat[Stilbene]{\includegraphics[width=3in]{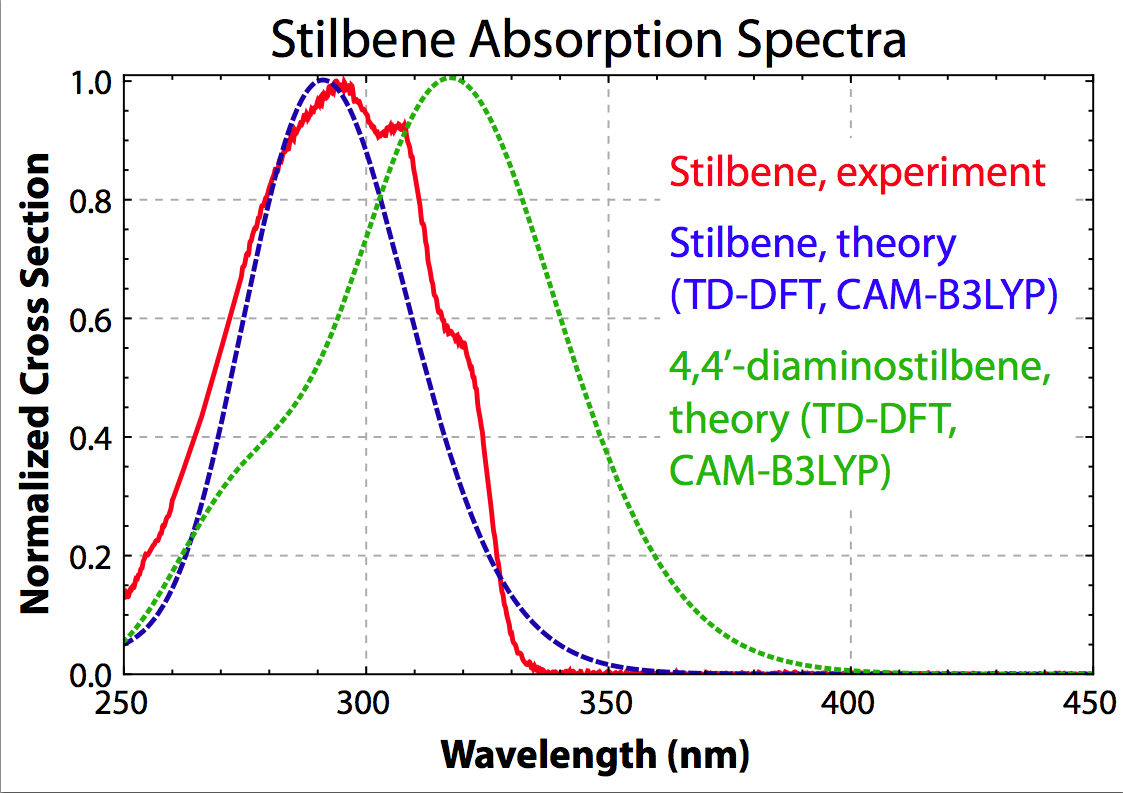}}
\caption{Comparison of spectra from TD-DFT calculations and experimental measurements using exchange/correlation functionals giving best match to data. These are identified in the figure legends.}
\label{stilbene_fluorescein_spectra}
\end{figure}

\begin{figure}
\centering{}
\subfloat[ ]{\includegraphics[width=3in]{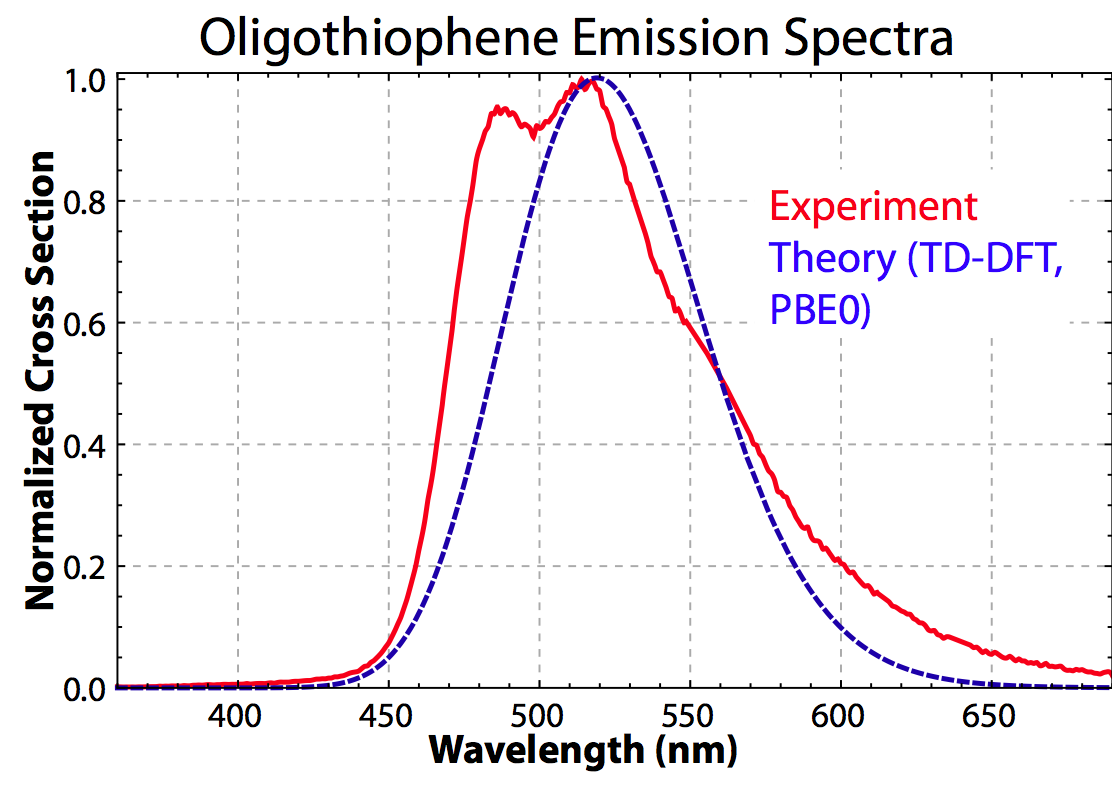}}
\qquad
\subfloat[ ]{\includegraphics[width=3in]{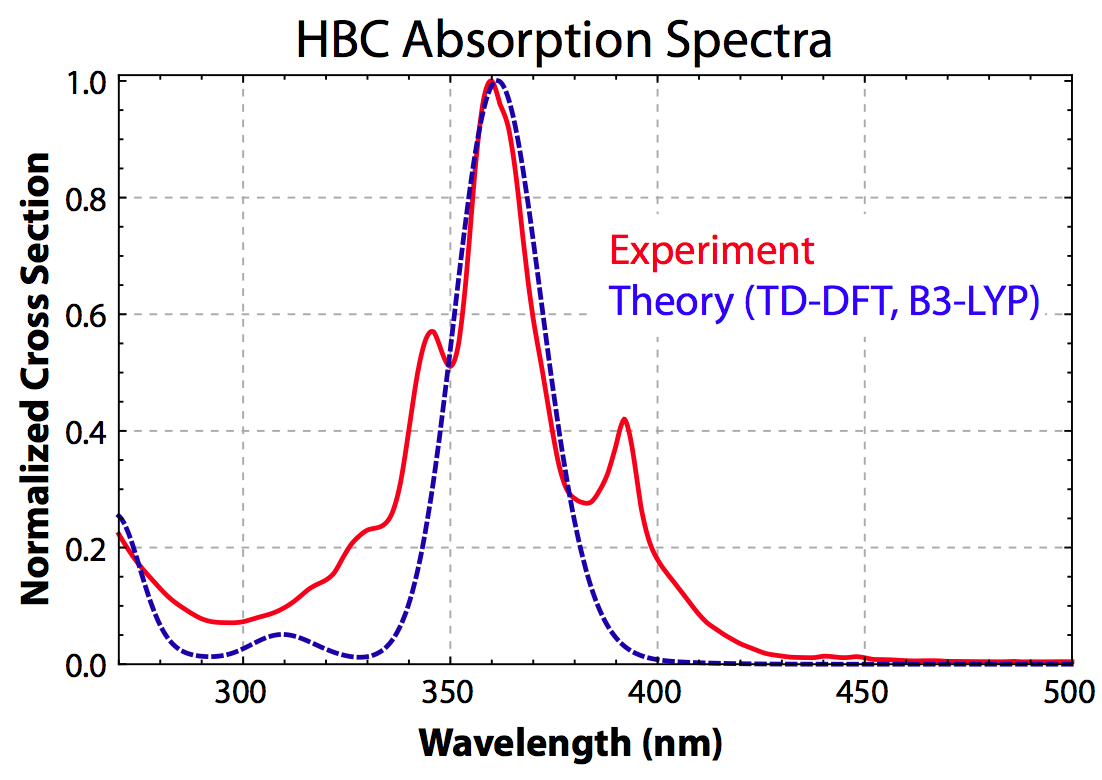}}
\caption{Comparison of spectra from TD-DFT calculations and experimental measurements using exchange/correlation functionals giving best match to data. These are identified in the figure legends.}
\label{HBC_OTP_Spectra}
\end{figure}

As shown in Fig. \ref{stilbene_fluorescein_spectra}, the CAM-B3LYP exchange/correlation functional  was determined to be best for stilbene while PBE0 gave the best match for fluorescein. The thermal broadenings used to generate the spectra were 250 meV for stilbene and 80 meV for fluorescein. In the case of stilbene, the experimental absorption data is for trans-stilbene rather than the trans-4,4$'$-diaminostilbene using for pooling calculations. A calculation was therefore performed for trans-stilbene in order to test the chemical accuracy of the exchange correlation functional with the functional chosen then used for all trans-4,4$'$-diaminostilbene calculations. The results are shown in Fig. \ref{stilbene_fluorescein_spectra}. 
%
The absorption and emission spectra for HBC and OTP, respectively, are shown in Fig. \ref{HBC_OTP_Spectra}. The exchange/correlation functional that gave the best chemical accuracy were B3LYP and PBE0 for HBC and OTP, respectively. The thermal broadenings used to generate the spectra were 100 meV for HBC and 150 meV for OTP.  The secondary peaks shown in the experimental data of HBC are most likely the result of a combination of phonon and aggregation effects that are not accounted for in the model calculations.
%
%

\section{RESULTS}

\subsection{Stilbene-Fluorescein}

\begin{figure}[h]
\centering{}
\subfloat[Intramolecular]{\includegraphics[width=3.2in]{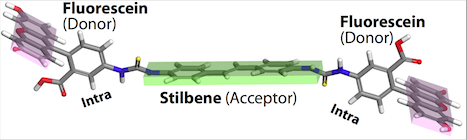}}
\qquad
\subfloat[Intermolecular]{\includegraphics[height=3.1in]{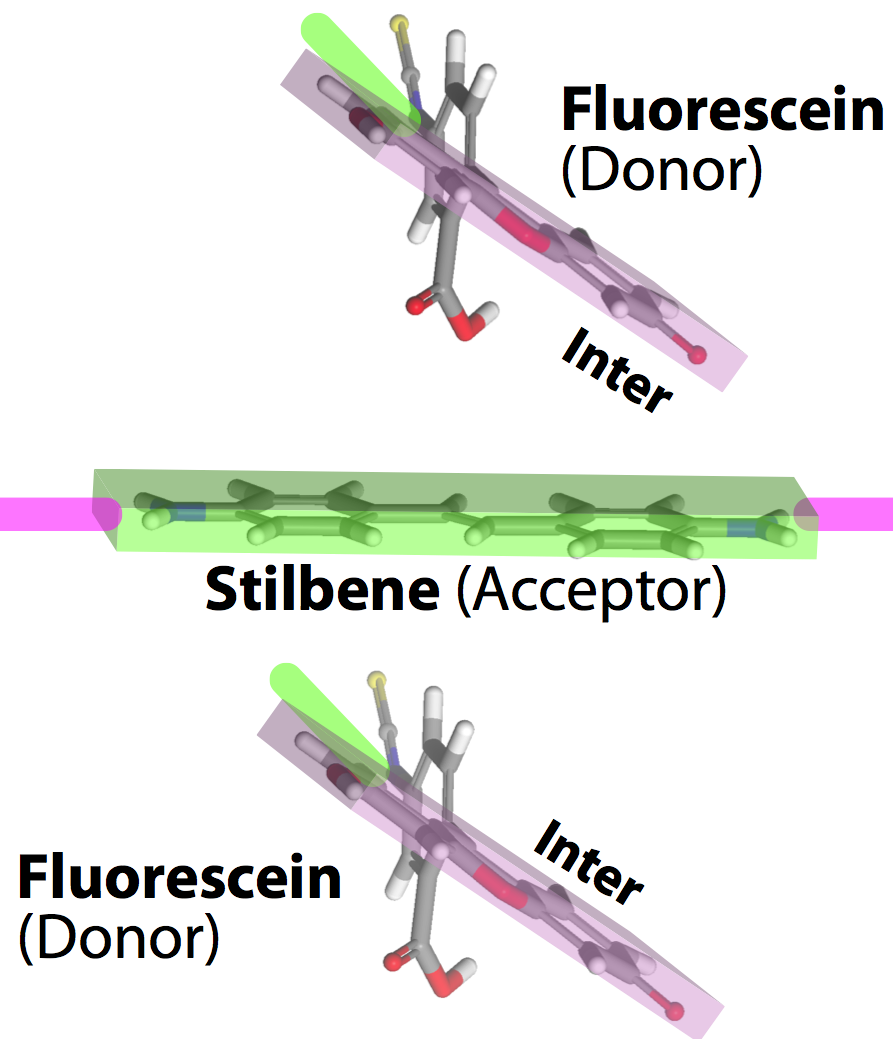}}
\caption{Molecular orientations of Stilbene-Fluorescein: (a) ground state geometry of a single molecule; (b) donor/acceptor orientations associated with optimized intermolecular pooling efficiency. S atoms are yellow, H atoms are white, C atoms are grey, O atoms are red, and N atoms are blue.  The green (Stilbene) and pink (Fluorescein) shading is a guide to the eye and is also used to indicate the location of moieties not explicitly accounted for in the computational analyses.}
\label{Stilbene_Fluorescein_Molecular_Orientations}
\end{figure}

Intermolecular and intramolecular dynamics were first considered for a single stilbene-fluorescein  molecule, shown in its ground state configuration in Fig. \ref{Fluorescein_Stilbene}(a). As a computational expedient, the molecule was divided into a 4,4$'$-diaminostilbene core and two fluorescein-5-isothiocyanate antennae with each moiety passivated and subjected to individual electronic structure analyses to obtain the required transition dipoles, electric-dipole coupling  tensor and two-photon absorption tensor. The interactions between the core and antennae were assumed to be well-characterized by simple dipole coupling allowing Eqs. (7)--(10) to be used to calculate the rates of energy pooling and the competing processes. The rates are shown in Table \ref{stilbene_fluorescein_rates}. The two orientations considered are displayed in Fig. \ref{Stilbene_Fluorescein_Molecular_Orientations}. In simplest terms it is a comparison of intermolecular energy pooling and intramolecular energy pooling.

\begin{table}[h]
\caption{Rates and efficiencies of stilbene-fluorescein system. Where the subscripts a and b refer to the geometric orientations shown in Fig. \ref{Stilbene_Fluorescein_Molecular_Orientations}}
\begin{centering}
\begin{tabular}{|c|c|c|c|c|}
\hline
 & $\Gamma_a (\mu s^{-1})$ & $\phi_a (\%)$ & $\Gamma_b (\mu s^{-1})$ & $\phi_b (\%)$\tabularnewline
\hline 
\hline 
PL & 8.1$\times$10$^1$ & 0.6 & 8.1$\times$10$^1$ & 0.4\tabularnewline
\hline
CET & 1.9$\times$10$^{0}$ & 0.0 & 3.2$\times$10$^3$ & 16.9\tabularnewline
\hline 
AET Total & 8.7$\times$10$^{-1}$ & 0.0 & 2.8$\times$10$^{1}$ & 0.1\tabularnewline
\hline 
ETU Total & 1.4$\times$10$^4$ & 99.4 & 1.6$\times$10$^4$ & 82.7\tabularnewline
\hline
\end{tabular}
\par\end{centering}
\label{stilbene_fluorescein_rates}
\end{table}

%
In the intramolecular configuration, we see that the efficiency of the energy pooling is relatively small. The main factor for this is the center to center distances between the three molecules. Conversely, if we consider intermolecular system we see that smaller distance favors the energy pooling mechanism. There is an additional orientation factor that does favor the intramolecular orientation, but it is small compared to the effect separation plays. The original experimental results did not quantify the efficiency of upconversion in such a way that a direct comparison could be made.\citep{stilbenefluorescein}

The possibility of triplet-triplet annihilation (TTA) was also considered. In the geometry of the first excited state it was determined that the energy gap between the singlet state and the nearest lying triplet state was only 20 meV. This makes the possibility of intersystem crossing extremely likely. Additionally, when relaxed in the triplet state it was found that the energy gap widened to 400 meV, which means that the intersystem crossing in uni-directional; once the triplet state is formed it is very unlikely for it to revert back into a singlet. Since the rate intersystem crossing can not be calculated within our theoretical setting, the efficiency of this process is not known.

\subsection{Hexabenzocoronene-Oligothiophene}

Hexabenzocoronene-oligothiophene (HBC-OTP) was studied next because the emission energy of the OTP is over half the absorption energy of the HBC making energy pooling possible. In addition, the hexagonal shape of the HBC allowed for six antennae rather than two increasing the probability of two antennae being excited simultaneously, an important requirement for energy pooling. Last, it has been observed that HBC tends to form collimated liquid crystalline phases that can be aligned along a given direction by the application of an external electric field. This effect is believed to give rise to high carrier mobility.\citep{NanHu} The geometry is shown in Fig. \ref{HBC_OTP}. Each component was assumed to be insulated and studied individually. The alkyl chains were methyl terminated for the HBC, but not for OTP due to the possible impact on the molecular geometry.

\begin{figure}
\centering{}
\subfloat[Single HBC-OTP molecule]{\includegraphics[width=3in]{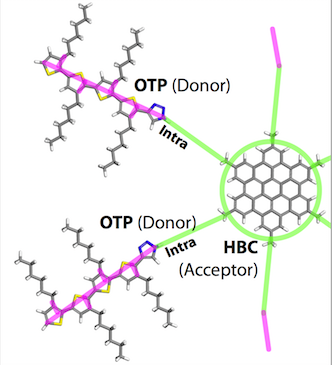}}
\qquad
\subfloat[Co-planar, inter-molecular geometry]{\includegraphics[width=3in]{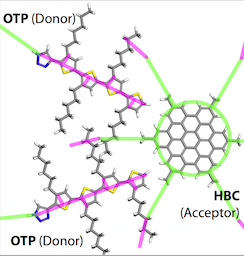}}
\qquad
\subfloat[Mixed inter-molecular, intramolecular geometry]{\includegraphics[width=3in]{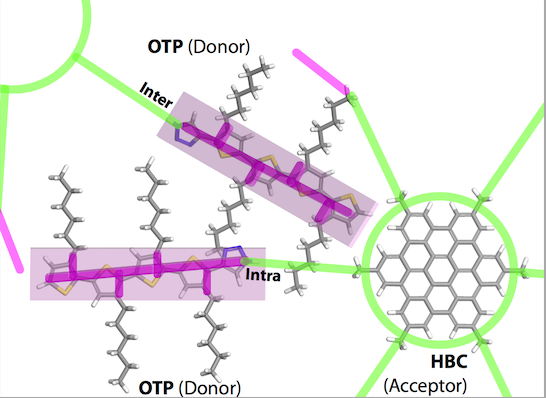}}
\qquad
\subfloat[Co-facial, intermolecular geometry]{\includegraphics[width=3in]{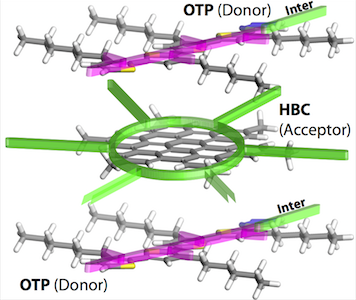}}
\caption{Four representative geometries for HBC-OTP. S atoms are yellow, H atoms are white, C atoms are grey, and N atoms are blue. The green (HBC) and pink (OTP) shading is a guide to the eye and is also used to indicate the location of moieties not explicitly accounted for in the computational analyses.}
\label{Molecular Orientations}
\end{figure}

Fig. \ref{Molecular Orientations} shows four possible orientations for HBC/OTP system. This includes the simplest intramolecular case of the antennae being nearest neighbors, an intermolecular configuration of two OTP from neighboring molecules interweaving between the antennae of the HBC, a hybrid between the intramolecular and intermolecular configurations, and an intermolecular configuration in which the OTP are situation on either side of the HBC at a distance of 7$\text{\AA}$.
 
Table \ref{hexabenzocoronene_oligothiophene} gives the results of the rates and efficiencies for each of the configurations of the HBC-OTP system. The efficiency of energy pooling in the intramolecular system is only about 0.4$\%$; high enough to be experimentally measurable but considerably less than desired. The main cause of this low rate is the relatively large distance between the geometric centers of the core and antennae. The fourth configuration is by far the most efficient in any case. This is due to the small center-to-center distance of the core and antennae in this configuration. If the distance factor is changed to match the distance in the first three configurations, the fourth configuration would be less efficient than the others due to the direction of the transition dipole moments.
\begin{table}
\caption{Rates and efficiencies of hexabenzocoronene-oligothiophene system. Where the subscripts a, b, c and d refer to the geometric orientations shown in Fig. \ref{Molecular Orientations}}
\begin{centering}
\begin{tabular}{|c|c|c|c|c|c|c|c|c|}
\hline
 & $\Gamma_a (\mu s^{-1})$ & $\phi_a (\%)$ & $\Gamma_b (\mu s^{-1})$ & $\phi_b (\%)$ & $\Gamma_c (\mu s^{-1})$ & $\phi_c (\%)$ & $\Gamma_d (\mu s^{-1})$ & $\phi_d (\%)$\tabularnewline
\hline 
\hline 
PL & 6.8$\times$10$^2$ & 5.5 & 6.8$\times$10$^2$ & 0.5 & 6.8$\times$10$^2$ & 1.6 & 6.8$\times$10$^2$ & 0.0\tabularnewline
\hline
CET & 5.5$\times$10$^1$ & 0.4 & 8.5$\times$10$^3$ & 6.6 & 6.7$\times$10$^2$ & 1.6 & 3.5$\times$10$^7$ & 99.0\tabularnewline
\hline 
AET (D-D') & 1.6$\times$10$^{-2}$  & 0.0 & 1.9$\times$10$^0$ & 0.0 & 2.0$\times$10$^{0}$ & 0.0 & 1.3$\times$10$^3$ & 0.0\tabularnewline
\hline
AET (D'-D) & 1.3$\times$10$^{-2}$  & 0.0 & 3.1$\times$10$^0$ & 0.0 & 8.7$\times$10$^{-2}$ & 0.0 & 1.3$\times$10$^3$ & 0.0\tabularnewline
\hline 
ETU (D-D') & 5.9$\times$10$^3$ & 48.2 & 5.9$\times$10$^4$ & 45.9 & 1.5$\times$10$^4$ & 33.7 & 1.8$\times$10$^5$ & 0.5\tabularnewline
\hline
ETU (D'-D) & 5.6$\times$10$^3$ & 45.8 & 6.1$\times$10$^4$ & 47.0 & 2.7$\times$10$^4$ & 63.2 & 1.8$\times$10$^5$ & 0.5\tabularnewline
\hline
\end{tabular}
\par\end{centering}
\label{hexabenzocoronene_oligothiophene}
\end{table}

%
\section{CONCLUSIONS}

Organic molecular assemblies tend to have strong electron-phonon coupling and a concomitant rapid internal conversion of excess exciton energy into heat. This is a primary obstacle in the design of organic upconversion materials. In such settings, energy pooling offers the prospect of efficiently up-converting excitonic energy by avoiding states in which hot excitons can lose energy to phonons. However, the conditions under which such relaxation rates become meaningful is not understood,  motivating the current investigation.  A combination of molecular quantum electrodynamics, perturbation theory, and electronic structure calculations was used to develop a simulator that calculates energy pooling rates along with other relaxation processes. These include photoluminescence and energy transfer upconversion but not triplet-triplet upconversion. Internal conversion was assumed to be much more rapid than any other relaxation dynamics. An important computational expedient was to assume that the interaction between donor and acceptor units is well-approximated by electric dipole coupling. This allowed rate expressions to be developed that can be populated with electronic structure data associated with isolated units--e.g. transition dipole elements for isolated donor molecules and isolated acceptor molecules.

Two specific example systems were chosen to exercise the methodology: stilbene-fluorescein and hexabenzocoronene-oligothiophene (HBC-OTP).  In the case of the stilbene-fluorescein, it was found that \underline{intra}-molecular energy pooling is measurable but with an efficiency less than 0.0$\%$. This can dramatically improved, however, to approximately 17$\%$ efficiency for geometry-optimized \underline{inter}-molecular dynamics. An examination of excited singlet and triplet energy levels, though, indicates that there is only small energy gap between the first singlet and the nearest triplet state. This leads us to conclude that triplet-triplet annihilation is also a reasonable interpretation for the upconversion observed in previous experiments, but the computational estimate of triplet-triplet annihilation is beyond the scope of the present investigation. For HBC-OTP, the efficiency of intramolecular energy pooling is estimated to be 0.4$\%$, primarily limited by the large spacing between donor and acceptor components. On the other hand, an optimized co-facial orientation that could be associated with either inter-molecular dynamics or OTP arms bent back onto the HBC yields pooling efficiencies of 99$\%$. Cooperative Energy Pooling was found to be much more efficient that Accretive Energy Pooling in all cases. We conclude that the future design efforts should focus on inter-molecular dynamics that employ the following design guidance:

\begin{itemize}
\item{Energy pooling has an extremely sensitive r$^{-12}$ distance dependence as compared with compared the r$^{-6}$ dependence of energy transfer upconversion. This suggests that donor-acceptor distance should be minimized to promote energy pooling. As separation between components is reduced, though, multipole effects will become increasingly important as will Dexter processes. }
\item{Both donor and acceptor components should be chosen to have a large Two-Photon Absorption (TPA) cross section. This is because both pooling rates correlate with the TPA tensor which are directly related to the TPA cross-section (see Methodology section). As a result, the significant body of nonlinear optics data on TPA materials can be fruitfully applied to exciton dynamics as well.}
\item{Donor molecules should be used that cannot sequentially absorb two photons--i.e. cannot carry out Energy Transfer upconversion (ETU). Along the same lines, energy transfer from acceptor to donor should be designed out. In both cases the intent is to exclude the possibility of hot excitons losing energy to heat.}
\item{All upconversion rates depend strongly on the relative orientation of donor and acceptor components. Our simulations show that it is possible to lower the coupling between the donors while maintaining a strong coupling between the donors and the acceptor. This would decrease the rate of ETU so as to reduce the subsequent generation of heat.}
\end{itemize}

%
\section{ACKNOWLEDGMENTS}
This research is supported by the NSF SOLAR Grant CHE-1125937.  All computations were carried out at the Golden Energy Computing Organization, Colorado School of Mines. We are pleased to acknowledge useful discussions with Prof. Gregory Scholes, Princeton University.

%

\newpage
\bibliography{Stilbene_Fluorescein_2}
\newpage

\end{document}